\documentclass[reviewcopy]{elsarticle}

\usepackage[reviewcopy]{adndt}
\usepackage{longtable}

\usepackage{mathptmx}

\usepackage{amsmath}

\usepackage{amssymb}

\biboptions{square,sort&compress}
\bibpunct[]{[}{]}{,}{n}{}{;}
\begin{document}

\begin{frontmatter}

\journal{Atomic Data and Nuclear Data Tables}


\title{Energy levels and radiative rates for Cr-like Cu VI and Zn VII}
  \author[One]{K. M. Aggarwal\corref{cor1}}
  \ead{K.Aggarwal@qub.ac.uk}
  \author[Two]{P. Bogdanovich\fnref{}} 
  \author[One]{F. P. Keenan\fnref{}}
  \author[Two]{R. Kisielius\fnref{}} 


  \cortext[cor1]{Corresponding author.}

  \address[One]{Astrophysics Research Centre, School of Mathematics and Physics, Queen's University Belfast,\\Belfast BT7 1NN,
Northern Ireland, UK}

  \address[Two]{Institute of Theoretical Physics and Astronomy, Vilnius University, A. Go\v{s}tauto 12, LT-01108 Vilnius, Lithuania}
\date{05.02.2016} 

\begin{abstract}  
Energy levels and radiative rates ($A$-values) for transitions in Cr-like Cu~VI 
and Zn~VII are reported. These data are determined in the quasi-relativistic 
approach (QR), by employing a very large {\em configuration interaction} (CI) 
expansion which is highly important for these ions. No radiative rates are 
available in the literature to compare with our results, but our calculated 
energies are in close agreement with those compiled by NIST and other available 
theoretical data, for a majority of the levels. The $A$-values (and 
resultant lifetimes) are listed for all significantly contributing E1, E2 and 
M1 radiative transitions among the energetically lowest 322 levels of each ion. \\

Received 11 February 2016; Accepted 11 Month 2016 \\


\end{abstract}

\end{frontmatter}




\newpage

\tableofcontents
\listofDtables
\listofDfigures
\vskip5pc


\section{Introduction}

Recently, we reported energy levels and radiative rates ($A$-values) for 
transitions in Cr-like Co~IV and Ni~V \cite{co4}, and here we provide similar results 
for two other ions, namely Cu~VI and Zn~VII.
Generally, ions with $Z \le 30$ are important for the study of astrophysical 
plasmas. However, they may also be relevant to research in  fusion plasmas, 
because often some of them are constituent of the reactor walls as impurities. 
There are several observed lines of both Cu and Zn ions  (see for example the 
CHIANTI database \cite{chianti1,chianti8} at 
${\tt {\verb+http://www.chiantidatabase.org+}}$), but we are 
not aware specifically of those for Cu~VI and Zn~VII. However,  many of their lines in the 
157--243 \AA ~wavelength range are listed in the {\em Atomic Line List (v2.04)}
of Peter van Hoof  (${\tt {\verb+http://www.pa.uky.edu/~peter/atomic/+}}$), 
because these are useful in the generation of synthetic spectra. Inspecting the
atomic and molecular database Stout \cite{stout}, one can see that there are only 
incomplete sets of level energies for both  Cu~VI and Zn~VII,
whereas radiative transition data are completely absent. Additionally, ions of Cu 
have been identified of particular interest to fusion studies \cite{iaea}, 
and hence their atomic data are required for the modelling of such plasmas.

Poppe et al. \cite{pop} were the first to measure energies for levels of the 
3d$^6$ $^5$D and 3d$^5$($^6$S)4p $^5$P$^{\rm o}$ configurations of Cu~VI from a 
laboratory spectrograph. Soon after, van Kleef et al. \cite{ams1} extended their 
work to measure additional lines of the 3d$^5$($^4$P)4p $^5$P$^{\rm o}$  and 
3d$^5$($^4$D)4p $^5$D$^{\rm o}$  multiplets, classifying 29 levels in total. 
Based on these identifications, van het Hof et al. \cite{ams2} calculated 
energies for the levels of the 3d$^6$ $^5$D term. They used 
{\em least square fitting} of Slater-Condon parameters, and their calculations 
were biased towards the known (observed or measured) results and iso-ionic, 
iso-electronic and iso-nuclear trends. As a result, their theoretical energies 
differed between $-4.6\%$ and $+3.3\%$ with the measurements. However, they 
also predicted an energy for one additional level, i.e. 3d$^5$($^4$S)4p 
$^3$P$^{\rm o}_0$. The work of  \cite{ams2} was further extended by Uylings and Raassen 
\cite{ams3} who predicted energies for an additional 6 levels of the 3d$^5$4p 
configuration. However, the most extensive experimental and theoretical work 
has been performed by Raaasen and van Kleef \cite{ams4} who identified 
most of the levels of the 3d$^6$, 3d$^5$4s and 3d$^5$4p configurations of Cu~VI. 
Their listed (and other) energies have been compiled and assessed by Sugar and 
Musgrove \cite{sm1}, and their recommended values are also available on the 
NIST (National Institute of Standards and Technology) website 
\cite{nist15}. However, there are a few gaps in the energy spectrum (including 
the $^1$S$_0$ level of the ground configuration 3d$^6$) -- see Section~\ref{ene} 
and Table~\ref{ene-cu6}. Furthermore, no $A$-values are available in the 
literature for transitions of Cu~VI, and therefore the  aim of the present
paper is to complete the spectrum by predicting energies for the missing levels, 
as well as to calculate the $A$-values. 

To date the most complete experimental investigation for the energy levels of Zn~VII has 
been performed by van het Hof et al. \cite{ams5}, whose results have been 
compiled, assessed and recommended by Sugar and Musgrove~\cite{sm2}, and are 
also available at the NIST website \cite{nist15}. However, many levels (in fact 
more than for  Cu~VI) are missing from this compilation (see Section~\ref{ene} 
and Table~\ref{ene-zn7}), and therefore we have performed our calculations to 
predict energies for the missing levels as well as to report the $A$-values, 
which  are not available in the literature. 

\section{Energy levels}
\label{ene}

In this work, as in our other recent studies, we have employed the 
quasi-relativistic approximation (QR), described in detail elsewhere \cite{tro08}, 
to determine level energies, radiative lifetimes ($\tau$) and various transition 
parameters among the levels of the ground configuration 3d$^6$ and two lowest 
excited configurations 3d$^5$4s and 3d$^5$4p in Cu~VI and Zn~VII. The calculations
performed here are similar to those for Co~IV and Ni~V  \cite{co4}, and as for those ions, 
 we have also made test calculations with the 
general-purpose relativistic atomic structure package ({\sc grasp}) and the 
flexible atomic code ({\sc fac}). However, energies obtained with both these 
codes show just as large discrepancies  with measurements, in magnitude and 
orderings, as shown in table~A of \cite{co4} for Co~IV. Therefore, we discuss 
our theoretical results generated only in the QR approximation.

The one-electron radial orbitals (RO) for the electrons of the investigated 
configurations 3d$^6$, 3d$^5$4s and 3d$^5$4p, and for the 4d and 4f electrons,
were determined from the quasi-relativistic Hartree-Fock equations described
in \cite{qr06,qr07}. In the QR  calculations, 
relativistic effects are included through the Breit-Pauli approximation 
adopted for the quasirelativistic approximation \cite{tro08}. To include
the correlation, all one- and two-electron promotions from the $3\ell$ and 
$4\ell$ electron shells of the investigated configurations are considered in a 
large CI wavefunction expansion. For electrons with $5 \geq n \geq 8$ and 
$\ell < n$, which describe virtual excitations, the transformed radial
orbitals (TRO) are employed \cite{tro08}. The admixed configurations generated 
in this way produce over $10^9$ configuration state functions (CSFs), which makes 
the further calculations of $A$-values intractable. Therefore, we select only 
those admixed configurations that have the largest contributions to the CI 
wavefunction (see, e.g. \cite{mbpt05}). Following these procedures, the resultant 
CI basis consists of 425 even  and 256 odd configurations of Cu~VI, producing 669\,075 
and 991\,598 CSFs, respectively. In the case of Zn~VII, we 
include 398 even and 241 odd  configurations generating 648\,845 and
 960\,686 CSFs, respectively. As one can see from these numbers, 
the admixed configurations produce a very large set of CSFs even after the 
reduction procedures, described in \cite{csf02}. Because of this, 
we must limit the number of the admixed configurations by increasing the selection 
criterion to $10^{-5}$, and consequently some important configurations are omitted. 
Therefore, we cannot produce a satisfactory agreement of our results with the available 
measured energy levels for Cu~VI and Zn~VII. To improve the 
accuracy of the calculated data, all Slater integrals describing the $LS$-dependent 
interactions are (slightly) reduced by $2.5\%$ for Cu~VI and by $2.3\%$ for 
Zn~VII.

The energies obtained in the QR approximation are  listed in Table~\ref{ene-cu6} 
for all 322 levels of the 3d$^6$, 3d$^5$4s and 3d$^5$4p configurations of Cu~VI, 
along with the NIST values. Among the lowest 34 levels of the 3d$^6$ ground 
configuration, the maximum discrepancy between theoretical and experimental 
energies (for any level) is much less than 1\%, except for the levels of 
the lowest term $^5$D where it is above $1\%$ and reaches a maximum of $1.6\%$
for the level 2 ($^5$D$_3$) -- but even in this case the absolute energy 
deviation is just 19\,cm$^{-1}$. This is caused by insufficient accuracy in the 
calculation of the spin-orbit interaction. The accuracy achieved here is better than that
achieved for Co~IV and Ni~V \cite{co4},  because 
of the improvements made in our calculations. Furthermore, the discrepancies between 
the theoretical and experimental energies are much lesser -- usually less than
$0.2\%$ and not worse than $0.3\%$ -- for the remaining levels belonging to the
excited configurations 3d$^5$4s and 3d$^5$4p. We note that energies for 
many levels are missing from the NIST compilation (or the literature) and 
therefore, we have predicted energies for such levels -- see e.g., 62--89 in 
Table~\ref{ene-cu6}. Since the maximum discrepancy with the measurements among the 
levels of the excited configurations is less than 0.3\%, it should be a robust measure of 
accuracy for our predicted energies. Finally, the energy orderings between 
theory and measurement are nearly the same, although there are a few minor 
differences for close-lying levels, such as 28--29 and 38--42. 

However, the $LSJ$ designations listed in the table are not always definitive, 
because we have performed just a formal identification based on the maximum
percentage contribution of CSF in the CI wavefunction expansion, and some levels 
are highly affected by CSFs mixing. For this reason their description using just 
a $LSJ$ notation is not definitive in all cases, and other level identification
scheme have to be applied instead of an $LS$ designation. All such levels are shown 
by a superscript  ``a" or ``b" -- see e.g., levels 214, 220 and 269. However, we 
 stress that this problem is not unique to our code or our QR approximation.
It is a general atomic structure issue and applies to all such large  
calculations, by any code. This is why a large number of the 
 3d$^5$4p levels have no identified $LS$ terms in the NIST 
database \cite{nist15}.

In Table~\ref{comp} we compare our calculated energies with the theoretical work 
of Raaasen and van Kleef~\cite{ams4} and Uylings and  Raassen~\cite{ams3} for 
 common levels, but only for those for which experimental results are not 
available -- see their tables III and IV.  For all these levels there are no 
appreciable discrepancies among the three independent calculations.

Our calculated energies for the levels of Zn~VII are listed in 
Table~\ref{ene-zn7} along with those of the NIST compilation, which are available 
for about 50\% of the levels but cover almost the entire energy range. As for the
 Cu~VI results in Table~\ref{ene-cu6}, the discrepancies  with our calculations are 
smaller than $1\%$ for the lowest 34 levels of the 3d$^6$ ground configuration, except for 
 levels 2--5, for  the same reason as for the ground configuration levels
of Cu~VI. However, the agreement with measurement is significantly better --
usually better than  $0.05\%$ and no worse than $0.1\%$ (in only a few cases) for the levels 
of the excited 3d$^5$4s and 3d$^5$4p configurations. This should be  
indicative of the accuracy of our predicted levels, missing from the NIST 
compilation. Since data for the 3d$^5$4s level energies in the NIST compilation
are  missing, we assume our calculated results can serve as benchmarks 
for these levels.

\section{Radiative rates}
\label{rad}

To our knowledge, $A$-values are not available in the literature for 
radiative transitions in Cu~VI and Zn~VII. Therefore, in Table~\ref{tran-cu6} 
we list transition energies ($\Delta E$, cm$^{-1}$), wavelengths ($\lambda$, 
\AA), emission radiative rates ($A$-values, s$^{-1}$), weighted oscillator 
strengths ($gf$, dimensionless), and transition line strengths ($S$-values in 
atomic units) for the E1 (electric dipole), E2 (electric quadrupole), and M1 
(magnetic dipole) transitions of Cu~VI among all 322 levels summarised in Table~\ref{ene-cu6}
belonging to the lowest 3 configurations 3d$^6$, 3d$^5$4s and 3d$^5$4p. 
Similar results for Zn~VII are presented in 
Table~\ref{tran-zn7}. Only those $A$-values are listed in Tables~\ref{tran-cu6} 
and \ref{tran-zn7} which are $\geq 10\%$ of the largest value for an emission 
transition probability $A$ from the upper level $j$. This means that very weak 
transitions are not included to save on space, as their impact on the
modelling of plasmas shall be negligible. Due to this selection, the $A$-values 
for magnetic  quadrupole (M2) and electric octupole (E3) transitions are not 
included in Tables ~\ref{tran-cu6} and \ref{tran-zn7}, although their data are calculated. 
However, all the level energies and the radiative transition 
parameters, such as transition wavelengths, transition probabilities, oscillator
strengths and  line strengths for the E1, E2 and M1 transitions 
determined in the QR approximation, along with electron-impact excitation 
data determined in the plane-wave Born approximation, are freely available in the ADAMANT 
database at Vilnius University ({\tt http://www.adamant.tfai.vu.lt/database}). 
 We note that the E3 and M2 transitions
are very weak (i.e. smaller than $1\%$ of the strongest transition),  and therefore
their transition parameters have no entries in this database. 

Since no other data for $A$-values are available in the literature, it is 
difficult to assess the accuracy of our calculations. However, since our 
calculated energies are reliable to better than 1\% for a majority of levels, 
the corresponding data for $A$-values are likely to be reasonably accurate. 
Based on conclusions given in  \cite{co4} and the assessment
of the radiative transition data determined in the QR approximation from our 
previous work, we are confident that the current  $A$-values are reliable and
can be adopted in plasma spectra modelling.

To allow the application of our data to the 
modelling of  absorption spectra,  in Tables~\ref{abs-cu6} and \ref{abs-zn7} we 
list $\lambda$ and $f$-values for all {\em comparatively strong} absorption 
E1 transitions, i.e. those having $f \geq 0.10$. We  note that not all
important absorption transitions are present in Tables~\ref{tran-cu6} and
\ref{tran-zn7}. These transitions do not normally vary much with differing CI 
expansion basis. However, the remaining weak(er) transitions may show large 
variations with other independent calculations, as they are more 
susceptible to changes arising from  differing sizes of the CI wavefunction extension 
and varying the calculation method. Also in Tables~\ref{abs-cu6} and \ref{abs-zn7} 
we present $\lambda$ and $f$-values for some weaker absorption lines originating 
from the lowest 5 levels of the ground configuration term 3d$^6 \,\,^5$D. These 
lines may be useful for modelling the absorption spectra of low-temperature plasmas, 
as all the lowest levels of the $^5$D term are close.

\section{Radiative lifetimes and Land\'{e} $g$-factors}

The radiative lifetime $\tau$ of a level $j$ is determined as 
1.0/$\sum_{i} A_{ji}$, where the sum is over all calculated radiative decay
channels with  $i<j$. As for the $A$-values, no prior  theoretical or 
experimental results are available for $\tau$ for the two ions discussed here. 
Therefore, the accuracy of our calculated  $\tau$ should be  no worse than that 
for the $A$-values. Mainly it depends on the strongest emission transitions for
a particular level, as the influence of (numerically more) weak(er) transitions in the above sum is
less important. For the convenience of future workers, in Tables~\ref{ene-cu6} 
and \ref{ene-zn7} we summarise our values of $\tau$ for all levels. 

Also listed in the tables are the Land\'{e} $g$-factors,
which show how the energy levels split in a magnetic field. These are dimensionless
coefficients describing the Zeeman effect for a particular $LSJ$ level. In the
case of a multi-term, multi-configuration wavefunction, the Land\'{e} $g$-factor is
expressed as: 
\begin{equation}
g = 1 + \sum_{CLS}{\alpha(CLSJ)^2 \frac{J(J+1)-L(L+1)+S(S+1)}{2J(J+1)}}.
\end{equation}
Here $g$ is the Land\'{e} $g$-factor, the sum is over all CSFs for that level,
$C$ represents the configuration, $LSJ$ are total moments of the level, and 
$\alpha(CLSJ)$ is the percentage contribution of a particular CSF for the level 
eigenfunction. Comparing the theoretical values of $g$ with experiment, one can 
assess the quality of a multi-reference wavefunction.

\section{Conclusions}

Energy levels and radiative rates for transitions in Cu~VI and Zn~VII have 
been determined in the QR approximation. For the calculations, a very large CI 
wavefunction expansion basis has been adopted which helped to reduce the 
discrepancies between theory and measurement for energy levels. For the Cu~VI
ion, based on comparisons with measured (and limited theoretical) results, our 
energy levels of the excited configurations are assessed to be accurate to 
better than $0.2\%$ for most levels. In the case of the ground 
configuration levels, the accuracy usually is no worse than $0.7\%$, only with 
data for the lowest 5 levels (term $^5$D) being somewhat less reliable,
but their absolute energy discrepancies do not exceed a few tens of cm$^{-1}$.
For the Zn~VII ion, these differences are even smaller, and do not exceed $0.5\%$ 
for the excited configuration levels. The accuracy of the ground configuration 
level energies is approximately the same as for Cu~VI.

Our calculated energies are listed for all 322 levels originating from the 
3d$^6$, 3d$^5$4s and 3d$^5$4p configurations,  cover a much larger range 
than other available theoretical or experimental results, and  is a complete set
of energy levels for these three configurations. This is important  as
there are no available data for the 3d$^5$4s configuration of Zn~VII, whereas
only a few level energies have been determined for Cu~VI to date.

Corresponding data for radiative transition rates have also been calculated and 
 listed for the E1, E2 and M1 emission transitions among these 322 
levels. However, due to the paucity of prior results no comparisons (and hence 
accuracy assessments) can be made. Nevertheless, the accuracy achieved in the 
determination of energy levels, as well as the good agreement of $A$-values 
for similar multi-electron systems demonstrated in our previous publications, 
indicate our $A$-values should be reliable, particularly for  
comparatively strong transitions. 

A significantly extended data set for the Cu~VI and Zn~VII energy levels, including
not only their identification but also the wavefunction percentage composition,
are freely available from our database ADAMANT. In this we present much more extensive results
on the radiative transition parameters, including weaker transitions with a 
selection criterium decreased by a factor of 10. Electron-impact excitation cross 
sections and rates are also tabulated in ADAMANT. We believe our present data 
will be useful for the modelling of plasmas as well as for further 
accuracy assessments.

\ack
The work at QUB has been partially funded  by AWE Aldermaston.

\vspace*{1.0 cm}

\renewcommand{\baselinestretch}{1.0}
\footnotesize
\begin{longtable}{@{\extracolsep\fill}rlllll@{}}
\caption{\label{comp}
         Comparison of energies (in cm$^{-1}$) for some levels of Cu~VI -- 
         see Table~\ref{ene-cu6} for all levels.}
\\
Index  & Configuration & Level &  RK81 & UR96  & Present \\
 \hline\\
\endfirsthead\\
\caption[]{(continued)}
Index  & Configuration/Level    & RK81 & UR96  & Present   \\
\hline\\
\endhead
34 & 3d$^6$                & $^1_0$S$_0$         & 150319  &	       & 149833  \\
115& 3d$^5(^4_3{\rm P})$4p & $^5$D$_1^{\rm o}$   & 391590  & 391991 & 392553  \\
116& 3d$^5(^4_3{\rm P})$4p & $^5$D$_0^{\rm o}$   & 392085  & 392060 & 392583  \\
265& 3d$^5(^2_5{\rm S})$4p & $^3$P$_0^{\rm o}$   & 438365  & 438485 & 438912  \\
304& 3d$^5(^2_3{\rm P})$4p & $^1$S$_0^{\rm o}$   & 487714  & 487136 & 486762  \\
315& 3d$^5(^2_1{\rm D})$4p & $^3$D$_2^{\rm o}$   & 505688  & 505800 & 505334  \\
322& 3d$^5(^2_1{\rm D})$4p & $^1$P$_1^{\rm o}$   & 517278  & 517241 & 517447  \\
  \\ \hline  											       								   					      
\end{longtable}

\begin{flushleft}                                                                               
                               
{\small
RK81: Raassen and van Kleef  \cite{ams4} \\
UR96: Uylings and Raassen \cite{ams3}   \\                                                                               
}                                                                                               
                               
\end{flushleft} 
\end{document}